# A Critique on Caratheodory Principle of the Second Law of Thermodynamics

P. Radhakrishnamurty
42, Shivakrupa 1st Floor, 3A Main, Manjunathanagar, 2nd Phase
Bangalore – 560010, India. e-mail:padyala1941@yahoo.com

**Abstract**

Caratheodory's axiomatic formulation of the second law is considered as one of the standard forms of formulation of the law. However, it was mired in advanced mathematics. The formulation was strongly criticized by Max Planck, for it was based on the analysis of adiabatic processes and adiabatic accessibility. A thermodynamic process is said to be possible only if it satisfies both first law and the second law. A process is said to violate the second law only when it satisfies the first law, and violate the second law. We show here that violation of Caratheodory principle of the second law violates the first law itself. In other words, adiabatic inaccessibility arises only as a consequence of violation of the first law. This we demonstrate by assuming, in contradiction to Caratheodory's principle, that the states considered adiabatically inaccessible as adiabatically accessible, and prove that the assumption leads to no paradox – thereby showing that the adiabatic inaccessibility arises as a result of violation of the first law.
______________________________________________________________________

## Introduction

Caratheodory Principle of Second Law of Thermodynamics is rated as one amongst the most standard forms of the statement of the law. Till the beginning of the last century, the development of this law was considered to be lacking mathematical rigor [1]. The required rigor was supplied by the mathematician, Caratheodory, at the instance of Max Born, in 1909. His formulation of the second law was axiomatic and highly mathematical. Even so, many works follow Caratheodory line of approach. This is more so the case with the more mathematically inclined ones. However, it is not as popular as the second law formulations of Kelvin and Clausius. It was better understood only after Born explained the formulation in clearer terms. The development of Caratheodory's axiomatic thermodynamics is presented recently by L. Pogliani and M. N. Berbaran-Santos [2].

Thermodynamics predicts whether a given process under given conditions is or is not possible. Thermodynamics declares a process to be impossible, if it violates either the first law or the second law or both. Therefore, for a process to be possible in thermodynamics sense it must satisfy both the first and the second law. Processes that satisfy the first law need not necessarily satisfy the second law; but the converse is not true. All the processes that satisfy the second law must necessarily satisfy the first law.

A process is said to violate the second law, only if it satisfies the first law and violates the second law. Therefore, a valid statement of the second law must satisfy the basic requirement that the processes it addresses must satisfy the first law of thermodynamics.

We demonstrate in this note that, violation of Caratheodory principle of second law of thermodynamics implies violation of the first law itself. Consequently, Caratheodory principle cannot be considered as a statement of the second law of thermodynamics.

## Caratheodory's principle of the second law

Caratheodory Principle of Second Law of Thermodynamics states: "*In every arbitrarily close neighborhood of a given initial state there exist states that cannot be approached arbitrarily closely by adiabatic processes*" [3].

Caratheodory Principle of the First Law of Thermodynamics [3], as applicable to an adiabatic process undergone by a closed system, is given by the expression:

$$\Delta U = W \qquad (1)$$

Where $U$ is the internal energy of the system (forces acting at a distance and those of capillary action are assumed to be absent and not included in the forces of interaction between the system and surroundings), $\Delta U$ is the change in $U$ suffered by the system due to the process undergone by the system. $W$ is the energy in the form of mechanical work interaction between the system and the surroundings due to the said process♠.

By definition, a closed system undergoing an adiabatic process can neither exchange matter nor heat with the surroundings. Therefore, in an adiabatic process a system can exchange energy only in the form of work interactions with the surroundings.

## Adiabatic inaccessibility arises due to violation of first law

Let us consider any two arbitrarily close neighboring equilibrium states of a closed system A, B that satisfy the Caratheodory requirements [3]. Then, one of these two states, say B, is not accessible from the other state A, by an adiabatic process. But State A is accessible from state B by an adiabatic process. Let $U_A$ and $U_B$ represent the corresponding internal energies of the system in the two states.

We now assume, in contradiction to the Caratheodory principle of the second law, that the two chosen equilibrium states are approachable from each other by an adiabatic process. This assumption assures us a cyclic adiabatic process. We then demonstrate that this assumption leads to results that satisfy the first and second laws of thermodynamics. More specifically the assumption does not lead to a violation of the second law. Therefore, Caratheodory's adiabatic inaccessibility must arise only as a result of a violation of the first law. This is because impossibility of adiabatic cyclic processes can arise only as a result of violation of the first law of thermodynamics.

Let the system undergo an adiabatic process from one equilibrium state to another equilibrium state, say from A to B. Let the energy of work interaction for the process be $W$ units. Noting that $U$ is a state function, the first law demands,

---

♠We note that the statement of the first law of thermodynamics as applicable to an arbitrary process of a closed system includes one more term, $Q$, on RHS of eq (1). $Q$ is the energy in the form of heat interaction the system suffers due to the process. Thus, the first law as applicable to an arbitrary process of a closed system is expressed as: $\Delta U = W + Q$.

$$(U_B - U_A) = \Delta U = W \tag{2}$$

State B is not reachable by an adiabatic process from state A, only if $\Delta U \neq W$ (note that neither the system nor the surroundings suffer entropy change for the adiabatic cycle ABA). In other words state B is not reachable from state A by an adiabatic process only if the first law is violated. Thus, our demonstration is complete in this case.

Interlude: An elaboration is in order, here. When a process satisfies the first law it implies that the process is possible in both forward and reverse directions. The second law however, restricts that possibility only for certain processes – the so called reversible processes. For irreversible processes, ***the second law denies the possibility of occurrence of a process in a certain direction***. For example, the second law denies the direction of the process of conversion of energy in the form of heat to energy in the form of work in a one-temperature (1-T) cyclic process; but not the process *per se,* for, it allows the possibility of the process in the opposite direction - the direction of conversion of energy in the form of work to energy in the form of heat in a 1-T cyclic process. As a second example, we may consider transfer of energy in the form of heat from a body at a given temperature to another body at a lower temperature. Second law denies the possibility of occurrence of this process in the reverse direction.

The crux of the second law lies in the fact that it helps us to predict the direction in which a process that satisfies the first law occurs under given conditions. It is the first law that denies the process *per se* – when the process corresponds to perpetual motion of the first kind – a process that produces energy as output with no input energy or more output energy than the input energy. To conclude, the first law can deny the possibility of a thermodynamic process, the second law can only deny the possibility of a cyclic process in a certain direction. If a process AB is impossible, then the cycle ABA that takes the system back to its original state, either in clockwise direction or anticlockwise direction, becomes impossible. The second law is incapable of denying the occurrence of a cyclic process that satisfies the first law, both in clockwise and anticlockwise directions. Consequently, adiabatic inaccessibility is a consequence of violation of the first law. (End of interlude).

We now consider the adiabatic process from state B to state A that brings the system back to its original state. The work interaction $W$ for this process is such that

$$(U_A - U_B) = -\Delta U = -W \tag{3}$$

If, $-\Delta U \neq -W$, the first law is violated and our demonstration is complete.

For the complete cycle, ABA (or BAB),

$$W - W = \Delta U - \Delta U = 0 \tag{4}$$

Therefore, arbitrarily chosen neighboring equilibrium states A and B of a system can always be connected by an adiabatic process, if the process does not violate the first law.

The impossibility of approachability of an arbitrarily close equilibrium state in the neighborhood of a given equilibrium state of a system by an adiabatic process in Caratheodory principle, therefore, can arise only if, $\Delta U \neq W$ for that adiabatic process. The impossibility of the said process thus implies the violation of the first law of thermodynamics itself. This completes our demonstration.

## Acknowledgement

I wish to express my sincere thanks to Karel Netocny of the Institute of Physics AS CR, Prague for useful comments on an earlier version of this article.